\documentclass[11pt]{article}
\usepackage{amsmath,amssymb,amsfonts}
\usepackage[english]{babel}
\makeatletter
\@addtoreset{equation}{section}
\makeatother

\def\nn{\nonumber} \def\bd{\begin{document}} \def\ed{\end{document}}
\def\ds{\documentstyle}
\let\bm=\bibitem
\newcommand{\be}{\begin{equation}}
\newcommand{\ee}{\end{equation}}
\newcommand{\bea}{\setlength\arraycolsep{2pt} \begin{eqnarray}}
\newcommand{\eea}{\end{eqnarray}}
\newcommand{\hoch}[1]{$\, ^{#1}$}
\def\p{\partial}
\title{\large {\bf Extremal Kerr/CFT correspondence of five-dimensional rotating (charged)
black holes with squashed horizons}}
\date{}

\author{Jun-Jin Peng\footnote{pengjjph@163.com} ,\ Shuang-Qing Wu \\ \\
\small \sl College of Physical Science and Technology, Central China Normal University,\\
\small Wuhan, Hubei 430079, People's Republic of China \\
}

\begin{document}

\maketitle
\vspace{20pt}

\begin{center}
\textbf{Abstract}
\end{center}
A new holographic duality named Kerr/CFT correspondence was recently proposed to derive
the statistical entropy of four-dimensional extremal Kerr black holes via identifying
the quantum states in the near-horizon region with those of two-dimensional conformal
field theory living on the boundary. In this paper, we apply this method to investigate
five-dimensional extremal Kerr and Cveti\v{c}-Youm black holes with squashed horizons
in different coordinates and find that the near-horizon geometries are not affected by
the squashing transformation. Our investigation shows that the microscopic entropies are
in agreement with those given by Bekenstein-Hawking formula. In addition, we have also
investigated thermodynamics of the general non-extremal Cveti\v{c}-Youm black holes with
squashed horizons.

%%%%%%%%%%%%%%%%%%%%%%%%%%%%%%%%%%%%%%%%%%%%%%%%%%%%%%%%%%%%%%%%%%%%%%%%%
\voffset=-.90pt
\vspace{40pt}

%%%%%%%%%%%%%%%%%%%%%%%
\section{Introduction}\label{one}
%%%%%%%%%%%%%%%%%%%%%%%

Recently, a new duality called as Kerr/CFT correspondence has been put forward in \cite{GHSS}
to evaluate the microscopic entropy of a four-dimensional extremal Kerr black hole. This
method is very similar to the early work of Brown and Henneaux \cite{BH86} except that the
$AdS_3$ background is replaced by the near-horizon geometry of an extremal Kerr black hole
previously obtained in Ref. \cite{BH99}. It indicates that the generators of diffeomorphism,
which preserves an appropriate boundary condition for the near-horizon geometry of an extremal
Kerr black hole, are found to form a copy of Virasoro algebra in the two-dimensional conformal
field theory (CFT). By virtue of the Cardy formula, the microscopic entropy that matches the
one macroscopically calculated by the Bekenstein-Hawking area law can be derived in terms
of a generalized temperature related to the Frolov-Thorne vacuum. Compared with the previous
derivation \cite{SV} of the microscopic entropy of a five-dimensional extremal rotating
black hole \cite{BMPV}, this method does not use string theory or supersymmetry, like the
work of \cite{AS}. It is worth noticing that this method is essentially different from the
Carlip's work \cite{SC} in which the boundary is also restricted to the horizon, and the
statistical entropy of a general non-extremal black hole is derived via studying the
relationship between black hole thermodynamics and the two-dimensional near-horizon CFT
\cite{SC,SNS,MIP}.

Inspired with the extremal Kerr/CFT correspondence proposed in \cite{GHSS}, the method has
been generalized to study the entropies of extremal black holes in a lot of theories such
as the Einstein theory, string theory, and supergravity theory, as well as those solutions
in diverse dimensions
\cite{HHKNT,LMP,AOT1,HMNS,CCLP,ITW,AOT2,LS1,AMG,LMPV,PW,CW,DAS,GG,WYW,WT,CaoMT,AnninosH}.
The Kerr/CFT correspondence has also been applied to derive the CFT entropy of extremal
black rings \cite{LS2} and that of NS5-branes \cite{YN}. In all these extensions, the
central charges from the vector and scalar fields were usually neglected, but it was proven
in \cite{CMN} that they indeed make no contribution to the central charges. Subsequently,
this method was applied \cite{KK} to the case of extremal black holes in Einstein-Gauss-Bonnet
gravity theory, where it was demonstrated that the CFT entropy need not always be in agreement
with that via the Wald formula. However, if one takes into consideration the higher-derivative
corrections to the Einstein-Hilbert Lagrangian \cite{ACOTT}, they may coincide with each
other. Furthermore, making use of the holographic renormalization group flow, the Kerr/CFT
correspondence was extended in \cite{KH} to a more general gauge/gravity correspondence
in the full bulk spacetime of extremal black holes. This correspondence has been checked
\cite{ITW,PW,CW,WYW} within the context of five-dimensional minimal supergravity. Recently
we \cite{PW} have examined its validity in a five-dimensional, extremal, rotating (charged)
black hole immersed in G\"{o}del universe.

Although the Kerr/CFT correspondence has been checked in a wide range, a lot of questions,
such as how this duality takes place and properties of the conserved charges, are still open.
Along these lines, progress has been made in dynamics of the near horizon geometry
\cite{DynaDRS,DynaAHMR,DynaBHSS,DynaBSS,DynaCL}, boundary conditions \cite{BounMTY,BounRa},
applicability of the Cardy formula \cite{CarJN} and covariance of the conserved charges
\cite{CharAMR}. Other approaches to realize this dual can be found in
\cite{ApproBJSS,ApproCL,ApproCSZ}.

On the other hand, squashed Kaluza-Klein black holes \cite{IM,SSY,BR,TW} in five dimensions
attracted a lot of recent interest due to their many different features from their un-squashed
counterparts. All of them have a squashed three-sphere as their horizon topology, while at
spatial infinity they possess the same asymptotic geometry, namely, a nontrivial $S^1$ bundle
with constant fiber over the two-sphere in a four-dimensional Minkowski spacetime. With the
presence of such an asymptotic structure, one should take into consideration the contribution
of the gravitational tension to black hole thermodynamics \cite{CCO,KI1,KI2}. What is more,
when both the rotation and charge parameters \cite{MIMT}, and/or an additional G\"{o}del
parameter \cite{TIMN,TI,MINT,SSW} appear, one has to modify the first law of the squashed
black hole thermodynamics \cite{SSW} by incorporating the effect of the dipole charge
\cite{CH}. Some other interesting properties of these squashed Kaluza-Klein black holes
can be found in \cite{SqKK1,SqKK2}.

Therefore, it is important to check whether the Kerr/CFT correspondence still holds true
or not under the squashing transformation \cite{IM}. In this paper, we shall apply the above
Kerr/CFT correspondence to explore the entropies of an extremal squashed Kerr black hole
\cite{TW} and its charged extension \cite{MIMT} --- an extremal squashed Cveti\v{c}-Youm
(CY) black hole in five dimensions. These black hole solutions are generated by performing
a squashing transformation to the five-dimensional Myers-Perry black hole \cite{MP} with
two equal rotation parameters and the CY black hole \cite{CY} by setting three charges equal.
They are included as special cases of the general squashed version \cite{TIMN,TI,MINT,SSW}
of the five-dimensional Einstein-Maxwell-Chern-Simons-G\"{o}del black hole solution \cite{Wu}
when the G\"{o}del parameter is set to zero. Remarkably, we find that after taking the
near-horizon limit to the extremal squashed Kerr and CY black holes, the squashing
transformation takes no effect on their near-horizon geometries, and the same still
holds true for the extremal squashed black holes embedded in the G\"{o}del universe.

Our purpose of this paper is to examine the Kerr/CFT correspondence for the squashed black
holes in different coordinates and to see the central charges how to vary under the change
of the angular coordinates. The rest part of this paper is organized as follows. In Sec.
\ref{two}, we shall utilize the Kerr/CFT correspondence to calculate the entropy of a
five-dimensional extremal squashed Kerr black hole in the coordinates where the initial
radial coordinate remains unchanged. In Sec. \ref{three}, to compare with the uncharged
case, we investigate the extremal squashed CY black hole in a different coordinate system
where the apparent singularity at spatial infinity is removed and the angular coordinates
are the standard $2\pi$-period azimuthal ones. In Sec. \ref{four}, we will present the full
thermodynamical properties of a general non-extremal squashed CY black hole and re-derive
the generalized CFT temperature associated with the Frolov-Thorne vacuum by making use of
the first law of black hole thermodynamics. A brief summary is given in the last section.

%%%%%%%%%%%%%%%%%%%%%%%%%%%%%%%%%%%%%%%%%%%%%%%%%%%%%%%%%%%%%%%
\section{The extremal squashed Kerr black hole and its CFT duals} \label{two}
%%%%%%%%%%%%%%%%%%%%%%%%%%%%%%%%%%%%%%%%%%%%%%%%%%%%%%%%%%%%%%%

In Ref. \cite{GHSS}, the microscopic entropy of a four-dimensional extremal Kerr black hole
was obtained by identifying its near-horizon quantum states with those of a two-dimensional
conformal field theory living on the infinite boundary of the black hole's near-horizon
geometry. Following this method, we shall firstly derive the entropy of a five-dimensional
extremal Kerr black hole with a squashed $S^3$ horizon in this section.

%%%%%%%%%%%%%%%%%%%%%%%%%%%%%%%%%%%%%%%%%%%%%%%%%%%%%%%%%%%%%%%%%%%%%%
\subsection{Near-horizon geometry of a squashed Kerr black hole}
%%%%%%%%%%%%%%%%%%%%%%%%%%%%%%%%%%%%%%%%%%%%%%%%%%%%%%%%%%%%%%%%%%%%%%

Making use of the squashing transformation \cite{IM} to the five-dimensional Myers-Perry
black hole with two equal angular momenta, one can obtain a new black hole solution \cite{TW}
that possesses a squashed horizon. This squashed Kerr black hole is characterized by three
parameters $(m, a, r_{\infty})$ that correspond to the mass, angular momentum and the size
of a $S^1$ fiber at spatial infinity, respectively. Its metric is given by
\bea
ds^2 &=& -\frac{\chi^2\hat{r}^2\Delta(\hat{r})}{B(\hat{r})} d\hat{t}^2
 +\frac{K^2(\hat{r})}{\Delta(\hat{r})}d\hat{r}^2
 +\frac{1}{4}\hat{r}^2K(\hat{r})\big(d\theta^2 +\sin^2\theta d\psi^2\big) \nn \\
&& +\frac{1}{4}B(\hat{r})\Big\{d\hat{\phi} +\cos\theta d\psi
 -\chi\big[\omega(\hat{r}) -\omega (r_{\infty})\big]d\hat{t}\Big\}^2 \, ,
\label{Kersmetr}
\eea
where
\begin{subequations}
\begin{align}
\Delta(\hat{r}) &= \frac{(\hat{r}^2 -m)^2 -m(m -2a^2)}{\hat{r}^4} \, , \\
 B(\hat{r}) &= \frac{\hat{r}^4 +2m a^2}{\hat{r}^2} \, , \qquad \qquad
 \omega(\hat{r}) = \frac{4m a}{\hat{r}^4 +2m a^2} \, , \\
K(\hat{r}) &= \frac{(r_{\infty}^2 -m)^2 -m(m
 -2a^2)}{(r_{\infty}^2 -\hat{r}^2)^2} \, , \qquad
\chi = \frac{\sqrt{r_{\infty}^4 +2m a^2}}{\sqrt{(r_{\infty}^2 -m)^2 -m(m -2a^2)}} \, .
\end{align}
\end{subequations}
In Eq. (\ref{Kersmetr}), the radial coordinate $\hat{r}$ and the Euler angles ($\theta$,
$\psi$, $\hat{\phi}$) take $0 < \hat{r} < r_{\infty}$ and ($0 < \theta < \pi$, $0 < \psi
< 2\pi$, $0 < \hat{\phi} < 4\pi$), respectively. The line element (\ref{Kersmetr}) depicts
a five-dimensional black hole with squashed $S^3$ horizons. It has the asymptotical structure
that describes a twisted $S^1$ bundle over a four-dimensional Minkowski space-time when
$\hat{r}\to r_{\infty}$. This can be seen more clearly by sending $\rho\to \infty$ after
performing a coordinate transformation $\rho = \rho_0 \hat{r}^2/(r_{\infty}^2 -\hat{r}^2)^2$,
where $\rho_0 = r_{\infty}\sqrt{\Delta(r_\infty)}/2$. Taking the $r_{\infty}\to \infty$ limit,
one recovers the five-dimensional Kerr metric with two equal angular momenta. The dual CFT
entropy of a five-dimensional extremal Kerr black hole with two different rotation parameters
was investigated in \cite{LMP}.

The Hawking temperature, the entropy and the angular velocities of the event horizon can be
computed as
\bea
&& T = \frac{\chi\hat{r}_+ \Delta^{\prime}(\hat{r}_+)}{4\pi K(\hat{r}_+)
 \sqrt{B(\hat{r}_+)}} \, , \qquad
 S = \frac{1}{2}\pi^2\hat{r}_+^2K(\hat{r}_+)\sqrt{B(\hat{r}_+)} \, , \\
&& \Omega_{\psi}(\hat{r}_+) = 0 \, , \qquad\qquad
 \Omega_{\hat{\phi}}(\hat{r}_+) = \chi\big[\omega(\hat{r}_+)
 -\omega(r_{\infty})\big] \, ,
\label{therqua}
\eea
where $\prime$ denotes the differentiation with respect to the coordinate $\hat{r}$, and
$\hat{r}_+$ is the outer horizon given by ${\hat{r}_+}^2 = m +\sqrt{m(m -2a^2)}$. To obtain
a regular black hole solution, here and in what follows, we shall assume that $m\geq 2a^2$,
$a > 0$, and $\hat{r}_+ \ll r_{\infty}$. In particular, when $m = 2a^2$, the black hole
(\ref{Kersmetr}) becomes extremal.

Since our aim is to derive the entropy of an extremal squashed Kerr black hole via the CFT
duality, we now try to explore its near-horizon geometry. To do so, we need to perform
the following coordinate transformations:
\be
\hat{r} = \sqrt{2}a(1 +\lambda r) \, , \qquad
\hat{t} =\frac{a(r_{\infty}^2 -2a^2)}{2\sqrt{r_{\infty}^4
 +4a^4}}\cdot\frac{t}{\lambda} \, , \qquad
\hat{\phi} =\phi +\frac{r_{\infty}^4 -4a^4}{2(r_{\infty}^4
 +4a^4)}\cdot\frac{t}{\lambda} \, .
\ee
After taking the $\lambda\to 0$ limit, the near-horizon geometry of an extremal squashed Kerr
black hole reads
\be
ds^2 = \frac{a^2}{2}\Big(-r^2dt^2 +\frac{dr^2}{r^2} +d\theta^2
 +\sin^2\theta d\psi^2 \Big) +a^2(d\phi +\cos\theta d\psi +rdt)^2 \, ,
\label{etrsKerr}
\ee
which describes a three-sphere bundle over the $AdS_2$ space. It is worth noting that the
near-horizon metric (\ref{etrsKerr}) of an extremal squashed Kerr black hole takes the same
form as that of an extremal Kerr black hole without performing the squashing transformation
since the squashing function $K\to 1$ at the near-horizon limit under the extremity condition
$m = 2a^2$. In other words, we find that the squashing transformation does not change the
near-horizon geometry of an extremal Kerr black hole.

%%%%%%%%%%%%%%%%%%%%%%%%%%%%%%%%%%%%%%%%%%%%%%%%%%%%%%%%%%%%%%%
\subsection{Central charges and microscopic CFT entropy}
%%%%%%%%%%%%%%%%%%%%%%%%%%%%%%%%%%%%%%%%%%%%%%%%%%%%%%%%%%%%%%%

Now, we pay our attention to calculate the central charges of the near-horizon metric
(\ref{etrsKerr}). Before doing this, it is important to impose the appropriate boundary
conditions at spatial infinity and find the asymptotical symmetry group that preserves
these boundary conditions. Since the five dimensional near-horizon metric (\ref{etrsKerr})
possesses two $U(1)$ isometris,
we can choose two boundary conditions. Using $h_{\mu\nu}$ to denote the metric deviation
from the near horizon geometry (\ref{etrsKerr}), one of the boundary conditions is given
by
\begin{align}
\left(
\begin{array}{ccccc}
h_{tt}=\mathcal{O}(r^2) \quad & h_{tr}=\mathcal{O}(\frac{1}{r^2}) \quad
& h_{t\theta}=\mathcal{O}(\frac{1}{r}) \quad & h_{t\phi}=\mathcal{O}(1) \quad
& h_{t\psi}=\mathcal{O}(r) \\
& h_{rr}=\mathcal{O}(\frac{1}{r^3}) \quad & h_{r\theta}=\mathcal{O}(\frac{1}{r^2}) \quad
& h_{r\phi}=\mathcal{O}(\frac{1}{r}) \quad & h_{r\psi}=\mathcal{O}(\frac{1}{r^2}) \\
& ~& h_{\theta\theta}=\mathcal{O}(\frac{1}{r}) \quad
& h_{\theta\phi} = \mathcal{O}(\frac{1}{r}) \quad
& h_{\theta\psi}=\mathcal{O}(\frac{1}{r}) \\
& ~& ~& h_{\phi\phi}=\mathcal{O}(1) \quad & h_{\phi\psi}=\mathcal{O}(1) \\
& ~& ~& ~& h_{\psi\psi}=\mathcal{O}(\frac{1}{r})
\end{array}
\right).
\label{SqKerrBC1}
\end{align}
The most general diffeomorphism that preserves such a boundary condition takes the
form
\bea
\zeta &=& \left[C +\mathcal{O}(r^{-3})\right ]\partial_t
+\left[-r\partial_\phi \epsilon(\phi) +\mathcal{O}(1) \right] \partial_r \nn \\
&&+\mathcal{O}(r^{-1}) \partial_\theta
+ \left[\epsilon(\phi) +\mathcal{O}(r^{-2})\right] \partial_\phi
+\mathcal{O}(r^{-2})\partial_\psi
\label{diffA} \, ,
\eea
where $C$ is a constant and $\epsilon(\phi)$ is an arbitrary function. If we exchange
$\phi$ and $\psi$ in Eqs. (\ref{SqKerrBC1}) and (\ref{diffA}), we obtain the other
boundary condition and the corresponding general diffeomorphism.

Defining $\epsilon(\phi) =-e^{-\mathrm{i}n\phi}$ and
$\epsilon(\psi) =-e^{-\mathrm{i}n\psi}$, where $n$ are integers,
the asymptotic symmetry group can be generated by a class of diffeomorphisms
\be
\zeta_{n}^{(1)} = -e^{-\mathrm{i}n\phi}\p_{\phi}
 -\mathrm{i}nre^{-\mathrm{i}n\phi}\p_r \, , \qquad
\zeta_{n}^{(2)} = -e^{-\mathrm{i}n\psi}\p_{\psi}
 -\mathrm{i}nre^{-\mathrm{i}n\psi}\p_r \, ,
\ee
which satisfy Virasoro algebras
\be
\mathrm{i}\big[\zeta_{l}^{(i)}, \zeta_{n}^{(i)}\big]
 = (l -n)\zeta_{l+n}^{(i)} \qquad (i = 1, 2) \, .
\ee
It is worth noting that there exists another diffeomorphism $\partial_t$ that preserves
the boundary condition (\ref{SqKerrBC1}) and commutes with $\zeta_{n}^{(i)}$. As shown
in \cite{GHSS}, we impose a supplemental boundary condition that the conserved charge
generating the diffeomorphism $\partial_t$ vanishes to eliminate excitations above the
extremity. Each diffeomorphism $\zeta_{l}^{(i)}$ is associated to a conserved charge
defined by \cite{CMN,BBC}
\be
Q_{\zeta_{n}^{(i)}} = \int_{\p\Sigma}k^{gr}_{\zeta_{n}^{(i)}}[h, g] \, ,
\ee
where $\p\Sigma$ is a spatial slice that extends to the infinity, $h_{\mu\nu} =
\mathcal{L}_{\zeta}g_{\mu\nu}$ denotes the deviation from the background metric
(\ref{etrsKerr}), and the 3-form $k^{gr}_{\zeta_{n}^{(i)}}[h, g]$ is given by
\bea
k^{gr}_{\zeta}[h, g] &=& -\frac{1}{96\pi}\sqrt{-g}\epsilon_{\alpha\beta\gamma\rho\sigma}
 \Big[\zeta^\rho\nabla^\sigma h -\zeta^\rho\nabla_\nu h^{\sigma\nu}
 +\zeta_\nu\nabla^\rho h^{\sigma\nu} +\frac{1}{2}h\nabla^\rho\zeta^\sigma \nn \\
&& -h^{\rho\nu}\nabla_\nu\zeta^\sigma +\frac{1}{2}h^{\rho\nu}(\nabla^\sigma\zeta_\nu
 +\nabla_\nu\zeta^\sigma)\Big]dx^\alpha\wedge dx^\beta\wedge dx^\gamma \, ,
\label{grcenchar}
\eea
in which, $\zeta = \zeta_{n}^{(i)}$. The Dirac brackets of the conserved charges corresponding
to the diffeomorphisms $\zeta_{l}^{(i)}$ and $\zeta_{n}^{(i)}$ yield a common form of the
Virasoro algebras with central terms
\be
\int_{\p\Sigma}k^{gr}_{\zeta_{l}^{(i)}}[\mathcal{L}_{\zeta_{n}^{(i)}}g, g]
 = -\frac{\mathrm{i}}{12}c_i(l^3 +\beta l)\delta_{l +n, 0} \, ,
\label{centerm}
\ee
where $c_i$ denote the central charges corresponding to the diffeomorphisms $\zeta_{l}^{(i)}$,
$\beta$ is a trivial constant since it can be absorbed by a shift in $Q_{\zeta_{0}^{(i)}}$, and
\be
\mathcal{L}_{\zeta_{n}^{(i)}}g_{\rho\sigma} = \zeta_{n}^{(i)\nu} \p_{\nu} g_{\rho\sigma}
 +g_{\nu\sigma}\p_{\rho} \zeta_{n}^{(i)\nu} +g_{\nu\rho}\p_{\sigma}\zeta_{n}^{(i)\nu}
\ee
is the Lie derivative of the background metric (\ref{etrsKerr}) with respect to the vector
field $\zeta_{n}^{(i)}$. For the background metric (\ref{etrsKerr}), the conserved charges
associated with the diffeomorphisms $\zeta_{l}^{(1)}$ and $\zeta_{l}^{(2)}$ can be computed
as
\be
Q_{\zeta_{l}^{(1)}} = -\mathrm{i}\pi a^3(l^3 +2l)\delta_{l +n, 0} \, , \qquad
Q_{\zeta_{l}^{(2)}} = 0 \, .
\label{crecent}
\ee
By comparison of Eq. (\ref{centerm}) with (\ref{crecent}), the central charges can be read
off as
\be
c_1 = 12\pi a^3 \, , \qquad c_2 = 0 \, .
\label{cencvalu}
\ee
Eq. (\ref{cencvalu}) shows that the central charge associated with the coordinate $\psi$
vanishes. This is attributed to the choice of the angle coordinates $\theta$, $\psi$ and
$\hat{\phi}$. In fact, the metric (\ref{Kersmetr}) describes a black hole with two equal
but opposite angular momenta. Our choice of the coordinates just makes the angular momentum
related to the coordinate $\psi$ disappear but the one corresponding to $\hat{\phi}$ become
double. If one adopts the coordinates given in \cite{PW}, one can find that $c_1$ and $c_2$
have the same but nonzero values. To see this, we will adopt this kind of coordinates to
calculate the CFT entropy of an extremal CY black holes with squashed horizons in the next
section. What is more, Eq. (\ref{cencvalu}) supports that the central charge got from an
extremal black hole strongly relies on the rotational $U(1)$ isometry.

After obtaining the central charges of an extremal squashed Kerr black hole (\ref{etrsKerr}),
in order to get its CFT entropy, we have to calculate the generalized temperature with respect
to the Frolov-Thorne vacuum. The so-called CFT temperature can be evaluated by
\be
T_F = -\lim_{\hat{r}_+\to r_e}\frac{T}{\Omega_{\hat{\phi}}(\hat{r}_+)
 -\Omega_{\hat{\phi}}(r_e)} = \frac{1}{2\pi} \, ,
\ee
where $r_e = \sqrt{m} = \sqrt{2}a$ is the degenerate horizon of an extremal squashed Kerr
black hole. In favor of the Cardy formula for the CFT entropy at the temperature $T_F$, we
can obtain the microscopic entropy of the extremal squashed Kerr black hole
\be
S_{CFT} = \frac{\pi^2}{3}c_1T_F = 2\pi^2 a^3 \, ,
\label{eKerrentro}
\ee
which precisely agrees with the Bekenstein-Hawking entropy derived from Eq. (\ref{therqua}).

%%%%%%%%%%%%%%%%%%%%%%%%%%%%%%%%%%%%%%%%%%%%%%%%%%%%%%%%%%%%%%%%%%%%%%%%%%%%%
\section{The extremal squashed Cveti\v{c}-Youm black hole and its dual CFTs} \label{three}
%%%%%%%%%%%%%%%%%%%%%%%%%%%%%%%%%%%%%%%%%%%%%%%%%%%%%%%%%%%%%%%%%%%%%%%%%%%%%

In this section, we shall utilize the analysis parallel to the previous section to investigate
the CFT entropy of an extremal squashed CY black hole \cite{MIMT} within a coordinate system
that is regular at spatial infinity and possesses the angular coordinates different from those
of an extremal squashed Kerr black hole. Thanks to the existence of an additional $U(1)$ gauge
field, one has to consider its effect on the central charges. However, an explicit calculation
shows that the contribution from the gauge field is zero. Besides, our calculations also prove
that the squashing transformation still does not affect the near-horizon geometry of an extremal
squashed CY black hole.

%%%%%%%%%%%%%%%%%%%%%%%%%%%%%%%%%%%%%%%%%%%%%%%%%%%%%%%%%%%%%%%%%%%%%%%%%%%%%%%%%%
\subsection{Squashed Cveti\v{c}-Youm black hole and its near-horizon geometry}
%%%%%%%%%%%%%%%%%%%%%%%%%%%%%%%%%%%%%%%%%%%%%%%%%%%%%%%%%%%%%%%%%%%%%%%%%%%%%%%%%%

The squashed CY black hole solution can be constructed by applying the squashing procedure
\cite{IM,TW} to a five-dimensional rotating charged black hole solution with three equal
$U(1)$ charges found by Cveti\v{c} and Youm \cite{CY} in string theory. The solution fulfils
the complete Einstein-Maxwell-Chern-Simons equations in $D = 5$ minimal supergravity theory,
whose bosonic part is described by the Lagrangian
\be
I = \frac{1}{16\pi}\int d^5x \Big[\sqrt{-g}\big(R -F_{\mu\nu}F^{\mu\nu}\big)
 -\frac{2}{3\sqrt{3}}\epsilon^{\mu\nu\alpha\beta\gamma}F_{\mu\nu}
 F_{\alpha\beta}\mathcal{A}_{\gamma}\Big] \, ,
\label{Lagran}
\ee
where $\epsilon^{\mu\nu\alpha\beta\gamma}$ is the Levi-Civita tensor density. The metric of
the squashed CY black hole and the gauge potential that solve the motion equations derived
from the action (\ref{Lagran}) takes the following forms
\bea
ds^2 &=& -\frac{\Delta(r)}{B(r)}d\tilde{t}^2 +\frac{K^2(r)}{\Delta(r)}dr^2
 +\frac{1}{4}r^2K(r)\big(d\theta^2 +\sin^2\theta d\hat{\psi}^2\big) \nn \\
&& +\frac{1}{4}B(r)\big[d\tilde{\phi} +\cos\theta d\hat{\psi}
 -F(r)d\tilde{t}\big]^2 \, , \label{sCYmetr} \\
A &=& \frac{\sqrt{3}q}{2r^2}\Big[d\tilde{t} -\frac{a}{2}\big(d\tilde{\phi}
 +\cos\theta d\hat{\psi}\big)\Big] \, ,
\label{gaugepo}
\eea
where
\begin{subequations}
\begin{align}
\Delta(r) &= \frac{(r^2 -m)^2 -(m -q)(m +q -2a^2)}{r^4} \, , \qquad
 K(r) = \frac{r_{\infty}^4\Delta(r_{\infty})}{(r_{\infty}^2 -r^2)^2} \, , \\
 B(r) &= r^2 +\frac{2(m -q)a^2}{r^2} -\frac{q^2a^2}{r^4} \, , \qquad
 F(r) = \frac{2a}{B(r)}\Big(\frac{2m -q}{r^2} -\frac{q^2}{r^4}\Big) \, ,
\end{align}
\end{subequations}
and the coordinates $(r, \theta, \tilde{\phi}, \hat{\psi})$ take the same ranges as those of
a squashed Kerr black hole (\ref{Kersmetr}). The parameters $(m, a, q, r_{\infty})$, which
correspond to the mass, the angular momenta, the charge, and the spatial infinity, respectively,
are required to satisfy $m > 0$, $m\geq q$, $m\geq 2a^2 -q$, and $0 < r_- \leq r_+ < r_{\infty}$
so that the metric can describe a regular black hole. The outer/inner horizons $r_{\pm}$ of the
squashed CY black hole are determined by
\be
r_{\pm}^2 = m\pm \sqrt{(m -q)(m +q -2a^2)} \, .
\ee
When $r_{\infty}\to \infty$, the metric (\ref{sCYmetr}) becomes that of the original CY
black hole, whose CFT entropies have been derived via the Kerr/CFT correspondence method
in \cite{CCLP}.

There exists an apparent singularity at $r = r_{\infty}$ in the coordinate system $(\tilde{t},
r, \theta, \tilde{\phi}, \hat{\psi})$. To remove this singularity and obtain the asymptotical
structure, we now perform the coordinate transformation
\be
 \hat{\rho} = \rho_0\frac{r^2}{r_{\infty}^2 -r^2} \, , \qquad
 \hat{\phi} = \tilde{\phi} -\Omega_\infty \tilde{t} \, , \qquad
 \hat{t} = \tilde{t}/\chi\, ,
\ee
where
\bea
\rho_0 &=& \frac{1}{2}r_{\infty}\sqrt{\Delta(r_{\infty})} \, , \qquad \qquad
\chi = \frac{\sqrt{B(r_{\infty})}}{2\rho_0} \, , \nn \\
\Omega_\infty &=& F(r_{\infty}) = \frac{2a[(2m -q)r_{\infty}^2
 -q^2]}{r_{\infty}^6 +2(m -q)a^2r_{\infty}^2 -q^2a^2} \, ,
\eea
and then recast the metric (\ref{sCYmetr}) and the gauge potential (\ref{gaugepo}) into
the forms
\bea
ds^2 &=& -\frac{\chi^2\rho_0^2V(\hat{\rho})}{\hat{\rho}(\hat{\rho} +\hat{\rho}_0)
 h(\hat{\rho})} d\hat{t}^2 +\frac{\hat{\rho}(\hat{\rho}
 +\hat{\rho}_0)}{V(\hat{\rho})}d\hat{\rho}^2 +\hat{\rho}(\hat{\rho}
 +\rho_0)\big(d\theta^2 +\sin^2\theta d\hat{\psi}^2\big) \nn \\
&& +h(\hat{\rho})\Big\{d\hat{\phi}+\cos\theta d\hat{\psi} +\chi\Big[\Omega_\infty
 -\frac{g(\hat{\rho})}{h(\hat{\rho})}\Big]d\hat{t} \Big\}^2 \, , \label{asymmetri} \\
A &=& \frac{q(\hat{\rho} +\rho_0)\sqrt{3}}{4\hat{\rho} r_{\infty}^2}\big[\chi(2
 -a\Omega_\infty)d\hat{t} -a\big(d\hat{\phi} +\cos\theta d\hat{\psi}\big)\big] \, ,
\label{asypote}
\eea
where the functions $V(\hat{\rho})$, $h(\hat{\rho})$, and $g(\hat{\rho})$ are given by
\begin{subequations}
\begin{align}
V(\hat{\rho}) &= \frac{[2(m -q)a^2 +q^2](\hat{\rho} +\rho_0)^2
 -2m r_{\infty}^2\hat{\rho}(\hat{\rho} +\rho_0)
 +r_{\infty}^4\hat{\rho}^2}{4r_{\infty}^2\rho_0^2} \, , \\
h(\hat{\rho}) &= \frac{r_{\infty}^6\hat{\rho}^3 +a^2(\hat{\rho}
 +\rho_0)^2[2(m -q)r_{\infty}^2\hat{\rho} -q^2(\hat{\rho}
 +\rho_0)]}{4r_{\infty}^4\hat{\rho}^2(\hat{\rho} +\rho_0)} \, , \\
g(\hat{\rho}) &= \frac{a(\hat{\rho} +\rho_0)[(2m-q)r_{\infty}^2\hat{\rho}
 -q^2(\hat{\rho} +\rho_0)]}{2r_{\infty}^4\rho^2} \, .
\end{align}
\end{subequations}
Clearly, the apparent singularity at $r = r_{\infty}$ of the metric (\ref{sCYmetr}) are
removed in the new coordinate system $(\hat{t}, \hat{\rho}, \theta, \hat{\phi}, \hat{\psi})$.
At spatial infinity $\hat{\rho}\to \infty$, the asymptotic metric becomes
\be
ds^2 = -d\hat{t}^2 +d\hat{\rho}^2 +\hat{\rho}^2\big(d\theta^2
 +\sin^2\theta d\hat{\psi}^2\big) +\frac{1}{4}B(r_\infty)(d\hat{\phi}
 +\cos\theta d\hat{\psi})^2 \, .
\label{refbm}
\ee
Therefore it is easy to see that the metric (\ref{asymmetri}) has the asymptotic geometry of
a twisted $S^1$ bundle over a four-dimensional Minkowski space-time when $\hat{\rho}\to \infty$.

We further make the angle coordinate transformations $\theta\to 2\theta$, $\hat{\phi}\to
\hat{\phi} +\hat{\psi}$, and $\hat{\psi}\to \hat{\phi} -\hat{\psi}$, and rewrite the metric
(\ref{asymmetri}) and the gauge potential (\ref{asypote}) as
\bea
ds^2 &=& -\frac{\chi^2\rho_0^2V(\hat{\rho})}{\hat{\rho}(\hat{\rho}
 +\hat{\rho}_0)h(\hat{\rho})}d\hat{t}^2 +\frac{\hat{\rho}(\hat{\rho}
 +\hat{\rho}_0)}{V(\hat{\rho})}d\hat{\rho}^2 \nn \\
&&+4\hat{\rho}(\hat{\rho}
 +\rho_0)\big[d\theta^2 +\sin^2\theta\cos^2\theta(d\hat{\phi} -d\hat{\psi})^2\big] \nn \\
&& +4h(\hat{\rho})\Big\{\cos^2\theta d\hat{\phi} +\sin^2\theta d\hat{\psi}
 +\frac{\chi}{2}\Big[\Omega_\infty -\frac{g(\hat{\rho})}{h(\hat{\rho})}\Big]d\hat{t}
 \Big\}^2 \, , \label{equametri} \\
A &=& \frac{q(\hat{\rho} +\rho_0)\sqrt{3}}{4\hat{\rho} r_{\infty}^2}\big[\chi(2
 -a\Omega_\infty)d\hat{t} -2a\big(\cos^2\theta d\hat{\phi}
 +\sin^2\theta d\hat{\psi}\big)\big] \, .
\label{equapote}
\eea
After doing these, the coordinates $\hat{\phi}$ and $\hat{\psi}$ in Eqs. (\ref{equametri})
and (\ref{equapote}) become the standard $2\pi$-period azimuthal ones. At the outer horizon,
the angular velocities with respect to them are equal and read
\be
\Omega_{\hat{\phi}}(\hat{\rho}_+) = \Omega_{\hat{\psi}}(\hat{\rho}_+)
 = \chi\big(\Omega_H-\Omega_\infty\big)/2 \, ,
\ee
where $\hat{\rho}_+ = \rho_0 r^2_+/(r_{\infty}^2 -r^2_+)$ is the outer horizon and $\Omega_H
= g(\hat{\rho}_+)/h(\hat{\rho}_+)$.

The Hawking temperature and the Bekenstein-Hawking entropy are
\begin{subequations}
\begin{align}
T &= \frac{\chi\rho_0V^\prime(\hat{\rho}_+)}{4\pi\hat{\rho}_+ (\hat{\rho}_+
 +\hat{\rho}_0)\sqrt{h(\hat{\rho}_+)}}
 = \frac{\chi(r_+^2 -r_-^2)}{2\pi r_+^2K(r_+)\sqrt{B(r_+)}} \, , \\
S &= 4\pi^2\hat{\rho}_+(\hat{\rho}_+ +\rho_0)\sqrt{h(\hat{\rho}_+)}
 = \frac{1}{2}\pi^2r_+^2K(r_+)\sqrt{B(r_+)} \, ,
\label{CYTEentr}
\end{align}
\end{subequations}
in which we have denoted $\prime = \p_{\hat{\rho}}$.

As before, our aim is to study the dual CFT entropies of an extremal squashed CY black hole.
The extremity conditions are $q = m$ or $q = 2a^2 -m$. When $q = m$, the extremal squashed
CY black hole becomes an extremal BMPV black hole with squashed horizon. In the following,
we shall derive the microscopic entropy of the extremal squashed CY black hole in these two
extremal cases. We will first discuss the extremal case $q = 2a^2 -m$, and leave the extremal
case $q = m$ for a separate subsection.

For the extremal case $q = 2a^2 -m$, we impose the supplement condition $a > 0$ and $m > a^2$.
In contrast with the case of an extremal squashed Kerr black hole, we shall implement our
task on the basis of the metric (\ref{equametri}) and the gauge potential (\ref{equapote}).
Under the extremity condition $q = 2a^2 -m$, we obtain their near-horizon forms
\bea
ds^2 &=& \frac{m}{4}\Big(-\rho^2 dt^2 +\frac{d\rho^2}{\rho^2}\Big)
 +m\big[d\theta^2 +\sin^2\theta \cos^2\theta(d\phi -d\psi)^2\big] \nn \\
&& +\frac{(m -a^2)(m +2a^2)^2}{m^2}(\cos^2\theta d\phi
 +\sin^2\theta d\psi +k\rho dt)^2 \, , \label{extreCYmetri} \\
A &=& -w\cos^2\theta \Big(d\phi +\frac{\sqrt{m-a^2}}{2a}\rho dt\Big)
 -w\sin^2\theta\Big(d\psi +\frac{\sqrt{m -a^2}}{2a}\rho dt\Big) \, ,
\label{extrepote}
\eea
after we make use of the coordinate transformations
\begin{subequations}
\begin{align}
\hat{\rho} &= \rho_e(1 +\lambda \rho) \, , \qquad
 \hat{t} =\frac{m\sqrt{h(\rho_e)}}{4\chi\rho_0\rho_e}\cdot\frac{t}{\lambda} \, , \\
\hat{\phi} &= \phi +\chi\big(\Omega_H -\Omega_\infty\big)\hat{t}/2 \, , \qquad
 \hat{\psi} = \psi +\chi\big(\Omega_H -\Omega_\infty\big)\hat{t}/2 \, ,
\end{align}
\end{subequations}
and send $\lambda\to 0$. In the above, $\rho_e = \rho_0m/(r^2_{\infty} -m)$, or $r_e =
\sqrt{m}$, is the location of the event horizon of an extremal squashed CY black hole,
while two constants $k$ and $w$ are given by
\be
k = \frac{a(3m -2a^2)}{2(m +2a^2)\sqrt{m -a^2}} \, , \qquad
w = \frac{\sqrt{3}a(2a^2 -m)}{2m} \, .
\ee
From Eqs. (\ref{extreCYmetri}) and (\ref{extrepote}), one can note that the squashing
transformation still does not affect the near-horizon geometry of an extremal squashed
CY black hole, since $K(r_e)\to 1$ when $r\to r_e$. This fact can be easily found from
the expression of the squashing function: $K(r) = (r_{\infty}^2 -r_-^2)(r_{\infty}^2
-r_+^2)/(r^2-r_{\infty}^2)^2$, which guarantees $K\to 1$ once the near-horizon limit
has been taken. Particularly, if we set $m = 2a^2$, then the near-horizon geometry of
an extremal squashed CY black hole (\ref{extreCYmetri}) reduces to that of an extremal
squashed Kerr black hole presented in the last section.

%%%%%%%%%%%%%%%%%%%%%%%%%%%%%%%%%%%%%%%%%%%%%%%%%%%%%%%%%%%%%%%
\subsection{Central charges and dual CFT entropies in the $q = 2a^2 -m$ case}
%%%%%%%%%%%%%%%%%%%%%%%%%%%%%%%%%%%%%%%%%%%%%%%%%%%%%%%%%%%%%%%

In this subsection, we will calculate the dual CFT entropies on the basis of the metric
(\ref{extreCYmetri}) and the gauge potential (\ref{extrepote}). As before, we first impose
the boundary conditions, which include the perturbations of the metric and the gauge field.
For the metric fluctuations around the background metric (\ref{extreCYmetri}), we choose
the same boundary conditions as those of the five dimensional squashed extremal Kerr black
hole in the previous section. For the gauge field, letting $a_\mu$ denote its perturbation,
we impose the boundary condition
\be
\left(a_t,a_r,a_\theta,a_\phi,a_\psi\right) \sim
\mathcal{O}(r,r^{-2},1,r^{-1},r^{-1}) \, .
\ee

The diffeomorphisms and the $U(1)$ gauge transformations that preserve these boundary
conditions are given by
\begin{subequations}
\begin{align}
\zeta_{n}^{(1)} &= -e^{-\mathrm{i}n\phi}\p_{\phi}
 -\mathrm{i}n\rho e^{-\mathrm{i}n\phi}\p_{\rho} \, , \qquad
 \Lambda^{(1)}_{n} = -w\cos^2\theta e^{-\mathrm{i}n\phi} \, , \\
\zeta_{n}^{(2)} &= -e^{-\mathrm{i}n\psi}\p_{\psi}
 -\mathrm{i}n \rho e^{-\mathrm{i}n\psi}\p_{\rho} \, , \qquad
 \Lambda^{(2)}_{n} = -w\sin^2\theta e^{-\mathrm{i}n\psi} \, ,
\end{align}
\end{subequations}
where $(n = 0, \pm 1, \pm 2, \cdots)$. These generators constitute the centerless Virasoro
algebras
\be
\mathrm{i}\big[(\zeta_{l}^{(i)}, \Lambda^{(i)}_{l}), (\zeta_{n}^{(i)},
 \Lambda^{(i)}_{n})\big] = (l -n)\big(\zeta_{l +n}^{(i)},
 \Lambda^{(i)}_{l +n}\big) \, , \qquad (i = 1, 2).
\ee
The combined generator $\big(\zeta_{l}^{(i)}, \Lambda^{(i)}_{l}\big) \equiv(\zeta, \Lambda)$
possesses an associated conserved charge $Q_{\zeta, \Lambda}$ defined by \cite{CMN,BBC}
\be
Q_{(\zeta,\Lambda)} = \int_{\p\Sigma}\Big(k^{gr}_{\zeta}[h, g]
 +k^{em}_{(\zeta,\Lambda)}[h,a; g,A] +k^{cs}_{(\zeta,\Lambda)}[h,a; g,A]\Big) \, .
\ee

From the Lagrangian (\ref{Lagran}), we note that $k^{gr}_{\zeta}$ is still given by Eq.
(\ref{grcenchar}), while $k^{em}_{(\zeta, \Lambda)}$ and $k^{cs}_{(\zeta, \Lambda)}$ read
\bea
k^{em}_{(\zeta,\Lambda)} &=& \frac{1}{48\pi}\sqrt{-g}\epsilon_{\alpha\beta\gamma\mu\nu}
 \Big[\Big(2h^{\mu\lambda} F_\lambda^{~\nu} -f^{\mu\nu} -\frac{1}{2}hF^{\mu\nu}\Big)
 (A_\rho\zeta^\rho +\Lambda) -F^{\mu\nu}a_\rho\zeta^\rho \nn \\
&& -2\zeta^\mu F^{\nu\lambda}a_\lambda -a^\mu g^{\nu\sigma}(\mathcal{L}_\zeta A_\sigma
 +\p_{\sigma} \Lambda)\Big]dx^\alpha\wedge dx^\beta\wedge dx^\gamma \, , \\
k^{cs}_{(\zeta,\Lambda)} &=& \frac{1}{2\sqrt{3}\pi}a_\alpha F_{\beta\gamma}
 dx^\alpha\wedge dx^\beta\wedge dx^\gamma \, ,
\eea
in which $a_\mu = \mathcal{L}_{\zeta}A_{\mu} +\p_{\mu}\Lambda$, and $f_{\mu\nu} =
\p_{\mu}a_{\nu} -\p_{\nu}a_\mu$. Taking into account the contribution from the $U(1)$
gauge transformations, the central term (\ref{centerm}) in the Virasoro algebras is
now modified to
\be
\int_{\p\Sigma}\Big(k^{gr}_{\zeta}[\mathcal{L}_{\tilde{\zeta}}g, g]
 +k^{em}_{(\zeta, \Lambda)}[\mathcal{L}_{\tilde{\zeta}}g,
 \mathcal{L}_{\tilde{\zeta}}A +d\tilde{\Lambda}; g,A]
 +k^{cs}_{(\zeta,\Lambda)}[\mathcal{L}_{\tilde{\zeta}}g,
 \mathcal{L}_{\tilde{\zeta}}A +d\tilde{\Lambda}; g,A]\Big) \, ,
\label{mocent}
\ee
where $(\tilde{\zeta}, \tilde{\Lambda})\equiv\big(\zeta_{n}^{(i)}, \Lambda^{(i)}_{n}\big)$.

Based upon the metric (\ref{extreCYmetri}) and the gauge potential (\ref{extrepote}), an
explicit calculation shows that $k^{em}_{(\zeta, \Lambda)}$ and $k^{cs}_{(\zeta,\Lambda)}$
make no contribution to the central term (\ref{mocent}). Therefore, we only need to consider
the contribution from the gravitational part $k^{gr}_{\zeta}$ as before and get
\be
\int_{\p\Sigma}k^{gr}_{\zeta_{l}^{(i)}}[\mathcal{L}_{\zeta_{n}^{(i)}}g, g]
 = -\frac{\mathrm{i}}{4}\pi k(m +2a^2)\sqrt{m -a^2} \Big[l^3 +\frac{2(m
 -a^2)(m +2a^2)^2}{m^3}l\Big]\delta_{l +n, 0} \, ,
\ee
from which the central charges $c_1$ and $c_2$, corresponding to the angle coordinates $\phi$
and $\psi$ respectively, can be read off as
\be
c_1 = c_2 = 3\pi k(m +2a^2)\sqrt{m -a^2} \, .
\ee
Note here that the central charges $c_1$ and $c_2$ are equal and nonzero. In contrast, if
one adopts another different angular coordinates from those of Eq. (\ref{extreCYmetri}),
one will find that one of the central charges of an extremal squashed CY black hole vanishes
but the other does not. As a matter of fact, when we calculate the central charges of an
extremal squashed CY black hole on the basis of the metric (\ref{asymmetri}) and the gauge
potential (\ref{asypote}), which have the same angle coordinates as those of an extremal
squashed Kerr metric (\ref{etrsKerr}), we also obtain $c^\prime_1 = 0$ and $c^\prime_2 =
2c_2$, where $c^\prime_1$ and $c^\prime_2$ denote the central charges with respect to the
angular coordinates $\hat{\phi}$ and $\hat{\psi}$ in Eq. (\ref{asymmetri}), respectively.

Bearing the above central charges in mind, we now have to calculate the dual CFT entropies
of the extremal squashed CY black hole. In terms of the generalized CFT temperature
\be
T_1 = T_2 = -\lim_{\hat{\rho}_+\to \rho_e}\frac{T}{\Omega_{\hat{\psi}}(\hat{\rho}_+)
 -\Omega_{\hat{\psi}}(\rho_e)} = \frac{1}{2\pi k} \, ,
\ee
we can directly derive the microscopic entropy via each of two copies of Virasoro algebras
by virtue of the Cardy formula and get
\be
S_{CFT} = \frac{\pi^2}{3}c_1 T_1 = \frac{\pi^2}{3}c_2 T_2
 = \frac{\pi^2}{2}(m +2a^2)\sqrt{m -a^2} \, ,
\label{eCYentr}
\ee
which matches the Bekenstein-Hawking entropy got from Eq. (\ref{CYTEentr}). In particular,
if $m = 2a^2$, namely $q = 0$, Eq. (\ref{eCYentr}) reproduces the dual CFT entropy
(\ref{eKerrentro}) of an extremal squashed Kerr black hole.

%%%%%%%%%%%%%%%%%%%%%%%%%%%%%%%%%%%%%%%%%%%%%%%%%%%%%%%%%%%%%%%%%%%%%%
\subsection{Dual CFT entropies under the extremity condition $q = m$}
%%%%%%%%%%%%%%%%%%%%%%%%%%%%%%%%%%%%%%%%%%%%%%%%%%%%%%%%%%%%%%%%%%%%%%

In this subsection, we shall derive the dual CFT entropies of an extremal squashed CY black
hole under the extremity condition $q = m$. In other words, we will take into consideration
the case of an extremal BMPV black hole with a squashed horizon. After taking the near-horizon
limit, the metric (\ref{equametri}) and the potential (\ref{equapote}) now become
\bea
ds^2 &=& \frac{m}{4}\Big(-\rho^2 dt^2 +\frac{d\rho^2}{\rho^2}\Big)
 +m\big[d\theta^2 +\sin^2\theta \cos^2\theta(d\phi -d\psi)^2\big] \nn \\
&& +(m -a^2)\Big[\cos^2\theta d\phi +\sin^2\theta d\psi
 -\frac{a\sqrt{m -a^2}}{2(m -a^2)}\rho dt\Big]^2 \, , \\
A &=& -\frac{\sqrt{3}}{2}a\Big[\cos^2\theta \Big(d\phi +\frac{\sqrt{m -a^2}}{2a}\rho
 dt\Big) +\sin^2\theta\Big(d\psi +\frac{\sqrt{m-a^2}}{2a}\rho dt\Big)\Big] \, . \nn \\
\eea

As before, the Chern-Simons term and the gauge field still make no contribution to the
central charges. Hence we just need to calculate the gravitational part and get
\be
\int_{\p\Sigma}k^{gr}_{\zeta_{l}^{(i)}}[\mathcal{L}_{\zeta_{n}^{(i)}}g, g]
 = \frac{\mathrm{i}}{8}\pi ma \Big[l^3 +\frac{2(m -a^2)}{m}l\Big]\delta_{l +n, 0} \, ,
\ee
from which the central charges $c_1$ and $c_2$, associated with the angle coordinates
$\phi$ and $\psi$ respectively, can be obtained as
\be
c_1 = c_2 = -\frac{3}{2}\pi ma \, .
\ee
The generalized CFT temperature now turns to be
\be
T_1 = T_2 = -\lim_{\hat{\rho}_+\to \rho_e}\frac{T}{\Omega_{\hat{\psi}}(\hat{\rho}_+)
 -\Omega_{\hat{\psi}}(\rho_e)} = -\frac{\sqrt{m -a^2}}{\pi a} \, .
\ee

With the help of the Cardy formula, the dual CFT entropies of an extremal squashed BMPV
black hole are presented as
\be
S_{CFT} = \frac{\pi^2}{3}c_1 T_1 = \frac{\pi^2}{3}c_2 T_2
 = \frac{\pi^2}{2}m\sqrt{m -a^2} \, .
\ee
This result agrees with the previous ones \cite{ITW,CW} obtained for an extremal,
supersymmetric BMPV black hole without making the squashing transformation.

%%%%%%%%%%%%%%%%%%%%%%%%%%%%%%%%%%%%%%%%%%%%%%%%%%%%%%%%%%%%%%%%%%%%%%%%%%%%%
\section{Thermodynamics of general non-extremal squashed Cveti\v{c}-Youm black
holes and the generalized CFT temperature} \label{four}
%%%%%%%%%%%%%%%%%%%%%%%%%%%%%%%%%%%%%%%%%%%%%%%%%%%%%%%%%%%%%%%%%%%%%%%%%%%%%

In this section, we shall discuss the thermodynamical properties of general non-extremal
squashed Cveti\v{c}-Youm black holes and calculate the conserved charges based upon the
metric (\ref{asymmetri}) and the gauge potential (\ref{asypote}). Making use of the first
law of black hole thermodynamics, we will derive the generalized CFT temperature with
respect to the Frolov-Thorne vacuum.

For the general non-extremal Cveti\v{c}-Youm with a squashed horizon, it is natural to choose
the asymptotic metric (\ref{refbm}) as the appropriate reference background solution; therefore,
following the work of \cite{CCO}, we can get the generalized Abbott-Deser mass and angular
momentum as follows:
\bea
M &=& \frac{1}{4}\pi\chi\big\{(r_\infty^4 -3q^2)r_\infty^4 +2m (r_\infty^4
 -2qa^2)r_\infty^2 -4(m -q)^2(r_\infty^2 +a^2)a^2 \nn \\
&&+2(m +q)q^2a^2 \big\}\times \left[r_\infty^6
 +2(m -q)a^2r_\infty^2 -q^2a^2 \right]^{-1} \, , \qquad \\
J &=& J_{\hat{\phi}} = \frac{\pi a[2(2m -q)r_\infty^6 -3q^2r_\infty^4
 +q^3a^2]}{8r_\infty^6} \, .
\eea
The angular momentum along the $\hat{\psi}$-direction is zero. However, if we perform our
calculations based upon the metric (\ref{equametri}), we will find that the angular momenta
$J_{\hat{\psi}} = J_{\hat{\phi}} = J$. The generalized Abbott-Deser mass and angular momentum
given above coincide with those obtained by the counterterm method \cite{MS}, which corresponds
to the results presented in \cite{SSW} when the G\"{o}del parameter $j = 0$.

Since the general non-extremal squashed CY black hole has the asymptotical geometry similar
to that of a squashed Kaluza-Klein black hole in five dimensions, one should take into account
the gravitational tension (per unit time), which can be computed via the counterterm method
and given by
\bea
\mathcal{T} &=& \chi\big\{2(r_\infty^2 -m)r_\infty^8 +[(2m -q)r_\infty^2
 -2q^2]^2a^2 -q^4a^2 +6(m -q)q^2a^4 \nn \\
&&-8(m -q)^2r_\infty^2a^4\big\}\times\big\{4[r_\infty^6
 +2(m -q)a^2r_\infty^2 -q^2a^2]\big\}^{-\frac{3}{2}} \, .
\eea
The size of the extra dimension with a $S^1$ circle at infinity is
\be
\mathcal{L} = \frac{2\pi\sqrt{r_\infty^6 +2(m -q)a^2r_\infty^2 -q^2a^2}}{r_\infty^2} \, .
\ee
The Komar mass is related to the Abbott-Deser or counterterm mass by $M_K = M
-\mathcal{T}\mathcal{L}/2$.

In addition to the global charge --- the electric charge, one more additional local charge
--- the dipole charge \cite{CH} enters into the first law of black hole thermodynamics also.
The electric charge $Q$ and the dipole charge $\mathcal{D}$ can be computed as
\be
Q = \frac{\sqrt{3}}{2}\pi q\, , \qquad
\mathcal{D} = -\frac{\sqrt{3}qa}{4r_\infty^2} \, .
\ee

The electro-static potential $\Phi$ can be obtained by
\be
\Phi = \xi^{\mu}A_{\mu}\big|_{r_+} -\xi^{\mu}A_{\mu}\big|_{r_\infty}
 = \frac{\sqrt{3}q\chi(r_+^4 -qa^2)(r_\infty^2 -r_+^2)}{2r_\infty[r_+^6
 +2(m -q)a^2r_+^2 -q^2a^2]} \, ,
\ee
where $\xi = \p_{\hat{t}} +\Omega\p_{\hat{\phi}}$ is the Killing vector normal to the
horizon, in which $\Omega = \chi(\Omega_H -\Omega_\infty)$ is the angular velocity
measured in a non-rotating frame relative to infinity.

Finally, the dipole potential $\Phi_\mathcal{D}$ can be computed as
\be
\Phi_\mathcal{D} = -\frac{\sqrt{3}\pi qa\chi}{r_\infty^4} \Big\{\frac{(r_\infty^4
 -qa^2)[(2m -q)r_+^2 -q^2]}{r_+^6 +2(m -q)a^2r_+^2 -q^2a^2} -(r_\infty^2 +q)\Big\} \, .
\ee

With all the thermodynamical quantities in hand, it can be verified that they fulfill
the differential and integral first laws of black hole thermodynamics
\bea
dM &=& TdS +\Omega dJ +\Phi dQ +\Phi_\mathcal{D} d\mathcal{D}
 +\mathcal{T}d\mathcal{L} \, , \label{dFLne} \\
2M &=& 3TS +3\Omega J +2\Phi Q +\Phi_\mathcal{D}\mathcal{D}
 +\mathcal{T}\mathcal{L} \, . \label{iFLne}
\eea

On the other hand, at the extremal limit, the differential and integral first laws become
\bea
dM &=& \Omega^{(e)} dJ +\Phi^{(e)}dQ +\Phi_\mathcal{D}^{(e)}d\mathcal{D}
 +\mathcal{T}d\mathcal{L} \, , \label{dFLe} \\
2M &=& 3\Omega^{(e)}J +2\Phi^{(e)}Q +\Phi_\mathcal{D}^{(e)}\mathcal{D}
 +\mathcal{T}\mathcal{L} \, , \label{iFLe}
\eea
where $\Omega^{(e)}$, $\Phi^{(e)}$, and $\Phi_\mathcal{D}^{(e)}$ can be obtained respectively
from $\Omega$, $\Phi$, and $\Phi_\mathcal{D}$ via the replacement $r_+ \to r_e = \sqrt{m}$.

Subtracting Eq. (\ref{dFLne}) from (\ref{dFLe}), and Eq. (\ref{iFLne}) from (\ref{iFLe}), we
arrive at two thermodynamical relations
\bea
dS &=& \frac{\Omega^{(e)} -\Omega}{T} dJ +\frac{\Phi^{(e)} -\Phi}{T} dQ
 +\frac{\Phi_\mathcal{D}^{(e)} -\Phi_\mathcal{D}}{T} d\mathcal{D}
 = 2\frac{dJ}{T_L} +\frac{dQ}{T_Q} +\frac{d\mathcal{D}}{T_{\mathcal{D}}} \, , \\
S &=& \frac{\Omega^{(e)} -\Omega}{T} J +2\frac{\Phi^{(e)}-\Phi}{3T} Q
 +\frac{\Phi_\mathcal{D}^{(e)} -\Phi_\mathcal{D}}{3T} \mathcal{D}
 = 2\frac{J}{T_L} +\frac{2Q}{3T_Q} +\frac{\mathcal{D}}{3T_{\mathcal{D}}} \, ,
\eea
which hold true in the sense of L' H\^{o}spital rule even at the non-extremal case. Taking
the extremal limit ($r_+ \to r_e$), we obtain the generalized CFT temperature
\be
T_L = T_1 = T_2 = \lim\limits_{r_+ \to r_e}\frac{2T}{\Omega^{(e)} -\Omega} \, ,
\ee
where the factor `2' is attributed to the fact we have adopted the metric (\ref{equametri})
and $J_{\hat{\psi}} = J_{\hat{\phi}} = J$. For an extremal squashed CY black hole at the
extremal case ($q = 2a^2 -m$), we get
\be
T_L = \frac{1}{2\pi k} \, ,
\label{TL1}
\ee
while for an extremal squashed BMPV black hole where $q = m$, we have
\be
T_L = -\frac{\sqrt{m -a^2}}{\pi a} \, .
\label{TL2}
\ee

To end up with our discussions, we shall derive the generalized CFT temperature in another
way by applying a similar thermodynamical relation that holds true for an extremal CY black
hole without making the squashing transformation. This is possible because the squashing
transformation does not affect the near-horizon geometry of an extremal squashed CY black
hole. In other words, the near-horizon geometry of an extremal squashed CY black hole is
identical to that of an extremal CY black hole.

For a general non-extremal CY black hole, its line element is given by Eq. (\ref{sCYmetr})
with $K(r) = 1$ or equivalently by setting $r_{\infty}\to \infty$. At the extremal case,
the relevant equation to determine the CFT temperature $T_L$ is
\be
dS = 2\frac{dJ}{T_L} +\frac{dQ}{T_Q} \, ,
\label{BMPVconstr}
\ee
where $S$, $J$ and $Q$ are the entropy, the angular momentum, and the electric charge of an
extremal CY black hole. Under the extremity condition ($q = 2a^2 -m$), they are
\be
S = \frac{\pi^2}{2}(m +2a^2)\sqrt{m -a^2} \, , \qquad
J = \frac{\pi}{4} a(3m -2a^2) \, , \qquad
Q = \frac{\sqrt{3}\pi}{2}(2a^2 -m) \, ,
\ee
where at the extremal BMPV case ($q = m$), they become
\be
S = \frac{\pi^2}{2}m\sqrt{m -a^2} \, , \qquad
J = \frac{\pi am}{4} \, , \qquad
Q = \frac{\sqrt{3}\pi m}{2} \, .
\ee
Substituting these quantities into Eq. (\ref{BMPVconstr}), we can re-derive the generalized
CFT temperature $T_L$ given by Eq. (\ref{TL1}) and (\ref{TL2}), respectively.

%%%%%%%%%%%%%%%%%%%%%%%%%%%%
\section{Conclusions} \label{CoRe}
%%%%%%%%%%%%%%%%%%%%%%%%%%%%

In this paper, we have applied the Kerr/CFT correspondence \cite{GHSS} to derive the
microscopic entropies of the five-dimensional extremal squashed Kerr and CY black holes.
After performing the near-horizon limit, we find that their near-horizon geometries are
described by a squashed three-spheres over $AdS_2$. Since under the extremity conditions,
the squashing function $K\to 1$ at the near-horizon limit, their near-horizon geometries
take the same ones as those of the extremal black holes without the squashing transformation.
When some suitable boundary conditions were imposed to the near-horizon geometries of
the extremal black holes, there exist a class of diffeomorphisms (and the $U(1)$ gauge
transformations for the charged case) that preserve these boundary conditions and generate
two copies of centerless Virasoro algebras. By calculating the Dirac brackets of the central
charges corresponding to the diffeomorphisms, one gets two copies of Virasoro algebras with
nonzero central terms, from which the central charges can be read off. By virtue of the
Cardy formula, together with the help of the generalized temperatures associated with the
Frolov-Thorne vacuum, we are able to derive the dual CFT entropies that precisely agree
with the Bekenstein-Hawking ones. It is worth noting that we have calculated the central
charges of the squashed Kerr and CY black holes in different angle coordinates. Our results
further support that the rotational $U(1)$ isometry plays a key role in determining the
central charge of the extremal black holes.

In addition, the thermodynamic properties of the general non-extremal squashed CY black
hole have been discussed. We have, for the first time, presented the explicit expression
for the dipole potential, in addition to the dipole charge and the gravitational tension,
that enters into the differential first law and the generalized Smarr relation. Based upon
the first law of black hole thermodynamics, we have re-derived the generalized CFT temperature.

In our forthcoming paper \cite{WP}, the complete thermodynamical properties of general
non-extremal Kerr-G\"{o}del and (charged) Einstein-Maxwell-Chern-Simons black holes with
squashed horizons in G\"{o}del universe have been studied. Besides, we will show that the
method of the extremal Kerr/CFT correspondence can be also applicable to reproduce the
Bekenstein-Hawking entropies in the extremal limit, although the computations become much
more involved. A paper about these aspects is in preparation.

\smallskip
%%%%%%%%%%%%%%%%%%%%%%%%%%
\textbf{Acknowledgments}:
%%%%%%%%%%%%%%%%%%%%%%%%%%
This work is partially supported by the Natural Science Foundation of China under Grant No.
10975058 and 10675051.

\end{document}